\documentstyle[12pt,worldsci]{article}
\input psfig.sty
\begin{document}
\setlength{\baselineskip}{2.6ex}
\def\nn{\noindent}
\def\Re{{\cal R \mskip-4mu \lower.1ex \hbox{\it e}\,}}
\def\Im{{\cal I \mskip-5mu \lower.1ex \hbox{\it m}\,}}
\def\ie{{\it i.e.}}
\def\eg{{\it e.g.}}
\def\etc{{\it etc}}
\def\etal{{\it et al.}}
\def\ibid{{\it ibid}.}
\def\sub#1{_{\lower.25ex\hbox{$\scriptstyle#1$}}}
\def\sul#1{_{\kern-.1em#1}}
\def\sll#1{_{\kern-.2em#1}}
\def\sbl#1{_{\kern-.1em\lower.25ex\hbox{$\scriptstyle#1$}}}
\def\ssb#1{_{\lower.25ex\hbox{$\scriptscriptstyle#1$}}}
\def\sbb#1{_{\lower.4ex\hbox{$\scriptstyle#1$}}}
\def\spr#1{^{\,#1}}
\def\spl#1{^{\!#1}}
\def\tev{\,{\rm TeV}}
\def\gev{\,{\rm GeV}}
\def\mev{\,{\rm MeV}}
\def\to{\rightarrow}
\def\dmix{\ifmmode D^0-\bar D^0 \else $D^0-\bar D^0$\fi}
\def\dm{\ifmmode \Delta m_D \else $\Delta m_D$\fi}
\def\rb{\ifmmode R_b\else $R_b$\fi}
\def\mh{\ifmmode m\sbl H \else $m\sbl H$\fi}
\def\mch{\ifmmode m_{H^\pm} \else $m_{H^\pm}$\fi}
\def\mt{\ifmmode m_t\else $m_t$\fi}
\def\mc{\ifmmode m_c\else $m_c$\fi}
\def\mz{\ifmmode M_Z\else $M_Z$\fi}
\def\mw{\ifmmode M_W\else $M_W$\fi}
\def\mws{\ifmmode M_W^2 \else $M_W^2$\fi}
\def\mhs{\ifmmode m_H^2 \else $m_H^2$\fi}
\def\mzs{\ifmmode M_Z^2 \else $M_Z^2$\fi}
\def\mts{\ifmmode m_t^2 \else $m_t^2$\fi}
\def\mcs{\ifmmode m_c^2 \else $m_c^2$\fi}
\def\mchs{\ifmmode m_{H^\pm}^2 \else $m_{H^\pm}^2$\fi}
\def\ztwo{\ifmmode Z_2\else $Z_2$\fi}
\def\zone{\ifmmode Z_1\else $Z_1$\fi}
\def\mtwo{\ifmmode M_2\else $M_2$\fi}
\def\mone{\ifmmode M_1\else $M_1$\fi}
\def\bsg{\ifmmode b\to s\gamma\else $b\to s\gamma$\fi}
\def\tb{\ifmmode \tan\beta \else $\tan\beta$\fi}
\def\xw{\ifmmode x\sub w\else $x\sub w$\fi}
\def\ch{\ifmmode H^\pm \else $H^\pm$\fi}
\def\lum{\ifmmode {\cal L}\else ${\cal L}$\fi}
\def\inpb{\ifmmode {\rm pb}^{-1}\else ${\rm pb}^{-1}$\fi}
\def\infb{\ifmmode {\rm fb}^{-1}\else ${\rm fb}^{-1}$\fi}
\def\epem{\ifmmode e^+e^-\else $e^+e^-$\fi}
\def\ppb{\ifmmode \bar pp\else $\bar pp$\fi}
\def\subw{_{\rm w}}
\def\half{\textstyle{{1\over 2}}}
\def\elli{\ell^{i}}
\def\ellj{\ell^{j}}
\def\ellk{\ell^{k}}
\newskip\zatskip \zatskip=0pt plus0pt minus0pt
\def\matth{\mathsurround=0pt}
\def\lsim{\mathrel{\mathpalette\atversim<}}
\def\gsim{\mathrel{\mathpalette\atversim>}}
\def\atversim#1#2{\lower0.7ex\vbox{\baselineskip\zatskip\lineskip\zatskip
  \lineskiplimit 0pt\ialign{$\matth#1\hfil##\hfil$\crcr#2\crcr\sim\crcr}}}
\def\undertext#1{$\underline{\smash{\vphantom{y}\hbox{#1}}}$}

\rightline{\vbox{\halign{&#\hfil\cr
&SLAC-PUB-95-6822\cr
&April 1995\cr}}}
\title{{\bf VIRTUAL EFFECTS OF PHYSICS BEYOND THE STANDARD MODEL }
\footnote{Work Supported by the Department of Energy,
Contract DE-AC03-76SF00515}
\footnote{Presented at the {\it 4th International Conference on Physics
Beyond the Standard Model}, Lake Tahoe, CA, December 13-18, 1994}}
\author{J.L.\ HEWETT\\
\vspace{0.3cm}
{\em Stanford Linear Accelerator Center, Stanford University, Stanford, CA
94309, USA}}
\maketitle

\begin{center}
\parbox{13.0cm}
{\begin{center} ABSTRACT \end{center}
{\small\hspace*{0.3cm}
\noindent We examine the indirect effects of new physics on a variety of
processes
in the $B$ system, such as the $Z\to b\bar b$ vertex, the decays $B\to X_s
\gamma$ and $B\to X_s\ell^+\ell^-$, and CP violation.
}}
\end{center}

\section{Introduction}

The investigation of virtual effects of new physics provides an important
opportunity to probe the presence of interactions
beyond the Standard Model (SM).
Various types of experiments may expose the existence of new physics,
including the search for direct production of new particles at high
energy accelerators.  Although this scenario has the advantage in that it
would yield the cleanest observation of new physics, it is limited by the
kinematic reach of colliders.  A complementary approach is offered
by examining the indirect effects of new interactions in higher order
processes, such as rare or forbidden decays and precision electroweak
measurements, and testing for deviations from SM predictions.
In fact, studies
of new loop induced couplings can provide a means of probing the detailed
structure of the SM at the level of radiative corrections where
Glashow-Iliopoulos-Maiani (GIM) cancellations are important.
In some cases the constraints on new degrees of
freedom via indirect effects surpass those obtainable from collider searches.
Given the large amount of high luminosity data which will become available
during the next decade, precision measurements and rare
processes will play a major role in the search for physics beyond the SM.
Here we highlight the importance of virtual effects in the $B$ system,
focusing on two-Higgs-doublet models, supersymmetry, and models
with anomalous couplings.

\section{The $Z\to b\bar b$ Vertex}

The SM continues to provide an excellent description of
precision electroweak data\cite{schaile}, especially in the light of the
discovery of the top-quark\cite{cdfdo} in the mass range predicted by this
data.  The only hint of a potential discrepancy is a mere $(2-2.5)\sigma$
deviation from SM expectations for the quantity $\rb\equiv\Gamma(Z\to b\bar b)/
\Gamma(Z\to {\rm hadrons})$.  A global fit to all LEP data gives the
result\cite{schaile} $\rb=0.2204\pm 0.0020$.  In this fit, the value of
\rb\ is highly correlated to the value of the corresponding quantity $R_c$,
which is measured to be $R_c=0.1606\pm 0.0095$.  If $R_c$ is fixed to its
SM value of 0.171, the LEP result for \rb\ becomes $0.2196\pm 0.0019$.
In the SM, \rb\ is sensitive to additional vertex corrections involving the
top-quark, while the remaining electroweak and QCD radiative corrections
largely cancel in the ratio.  These additional vertex corrections
suppress $\Gamma(Z\to b\bar b)$ by an amount which is approximately
quadratic in the top-quark mass and hence reduce
the value of \rb\ for the measured range of $m_t$.
This can be seen explicitly in Fig. 1(a) from Grant\cite{grant},
where the solid curves compare \rb\ in the SM (taking $m_h=100\gev$)
with the corresponding quantity $R_d$
which is not affected by this top-quark vertex correction.
Using ZFITTER4.9\cite{zfit} we find
$\rb=0.2157$, taking $m_t=175\gev$ (and $m_h=300\gev$, $\alpha_s=0.125$).

We first consider the effects of an enlarged Higgs sector on \rb.
We examine a two-Higgs-doublet model (denoted\cite{hhg} as Model II) which
naturally avoids tree-level flavor-changing neutral currents, and
where the second doublet, $\phi_2$, gives mass to the up-type quarks, while the
down-type quarks and charged leptons receive their mass from $\phi_1$. Each
doublet obtains a vacuum expectation value (vev) $v_i$, subject to the
constraint that $v_1^2+v_2^2=v^2$, where $v$ is the usual vev present in the
SM.  The fermionic couplings of the five physical Higgs bosons are
summarized in Ref. 5 and are dependent upon the fermion mass,
$\tb\equiv v_2/v_1$, and the neutral scalar mixing angle $\alpha$.
Two additional classes of $Zb\bar b$ vertex corrections are present in this
model\cite{grant,zbbsusy}; (i) the charged Higgs boson \ch\ being
exchanged together with the t-quark, and (ii) the neutral scalar $h^0, H^0$
and pseudoscalar $A^0$ Higgs bosons exchanged with the b-quark.
The diagrams involving the \ch\ exchange yield negative contributions to
$Z\to b\bar b$ and grow as $m_t$ increases,  thus further suppressing
this width.   The contributions from the neutral Higgs exchange can have
either sign, depending on the values of the scalar masses
(this correction is positive if
$h^0$ and $A^0$ have small degenerate masses), and become
important for large values of \tb.  This scenario is summarized\cite{grant}
by the dashed curves in Fig. 1(a), where the upper (lower) curves correspond
to $\tb=70(1)$ with $m_{h^0,A^0}=50\gev,\ m_{H^0}=875\gev$, $\mch=422\gev$,
and $\alpha=\pi/2$.  We see that it is possible to accommodate the data in
this model for very specific values of the parameters.

The (s)particles present in supersymmetric theories also
contribute to the $Zb\bar b$ vertex correction.  In addition to the SM and
Model II charged and neutral Higgs corrections discussed above, there are
further contributions from (i) top-squark-chargino and (ii) b-squark-neutralino
exchange.  These additional corrections have been examined by several
authors\cite{zbbsusy,wkk}, and have been found to be sizeable in some regions
of the parameter space.  In particular, the light $\tilde t_1$-chargino loops
can give large corrections, while the heavy
$\tilde t_2$-chargino and $\tilde b$-neutralino corrections decouple.
Wells \etal\cite{wkk}, have performed a phenomenological analysis of these
supersymmetric corrections and established that consistency at the $1\sigma$
level with the LEP data on \rb, together with the measured value of $m_t$,
requires light sparticles.  Specifically, they find
the constraint $m_{\tilde\chi_1^\pm}<85\gev$ and $m_{\tilde t_1}<100\gev$
with min$(\tilde\chi_1^\pm,\tilde t_1)<65\gev$.  This is shown explicitly in
Fig. 1(b) from this reference.

Anomalous $WWZ$ interactions would impact the top-quark corrections to the
$Zb\bar b$ vertex.
The tri-linear gauge boson vertex can be probed by
looking for deviations from the SM in tree-level processes such as
$\epem\to W^+W^-$, or in
loop order processes, for example the $g-2$ of the muon.  In the latter case,
cutoffs must be used in order to regulate the divergent loop integrals
and can introduce errors by attributing a physical significance to the
cutoff\cite{burgess}.
The CP-conserving interaction Lagrangian for $WWV$ interactions can be
written as\cite{dieter}
\begin{eqnarray}
{\cal L}_{WWV}& = & ig_{WWV}\left[ \left( W^\dagger_{\mu\nu}W^\mu
V^\nu-W^\dagger_\mu V_\nu W^{\mu\nu}\right) +\kappa_V W^\dagger_\mu W_\nu
V^{\mu\nu}+{\lambda_V\over\mws}W^\dagger_{\lambda\mu}W^\mu_\nu V^{\nu\lambda}
\right. \nonumber \\
& & \quad\quad\quad\quad \left. -ig_5^V\epsilon^{\mu\nu\lambda\rho}\left(
W^\dagger_\mu\partial_\lambda W_\nu-W_\nu\partial_\lambda W^\dagger_\mu\right)
V_\rho \right] \,,
\end{eqnarray}
where $V_{\mu\nu}=\partial_\mu V_\nu-\partial_\nu V_\mu$, $g_{WWV}=gc_w
(e)$ for $V_\mu=Z_\mu (A_\mu)$,
and the parameters ($\Delta\kappa_V\equiv\kappa_V-1$) take on
the values $\Delta\kappa_V, \lambda_V, g_5^V=0$ in the SM.
Eboli \etal\cite{eboli}, have examined the affect of
these anomalous interactions on \rb\ and derived the $95\%$ C.L. bounds
(for $m_t=175\gev$), $-1.2<\Delta\kappa_Z<-0.091,\
-6.0<\lambda_Z<-0.46,\ -1.9<g_5^Z<-0.14$, assuming that only one parameter
is non-zero at a time, and setting the cutoff scale $\Lambda=1$ TeV.  Negative
values of these parameters yield positive shifts in \rb\ and hence are
favored.  Note that these parameters must be unnaturally large in order to
accommodate the data.

The existence of anomalous couplings between the b-quark and
the $Z$ boson could cause a significant shift\cite{tgr} in the value of \rb.
The lowest dimensional non-renormalizable
operators which can be added to the SM take the form of either
electric or magnetic dipole form factors.  Defining $\kappa$ and $\tilde\kappa$
as the real parts of the magnetic and electric dipole form factors,
respectively, (evaluated at $q^2=M_Z^2$) the interaction Lagrangian is
\begin{equation}
{\cal L}={g\over 2c_w}\bar b\left[ \gamma_\mu(v_b-a_b\gamma_5) +
{i\over 2m_b}\sigma_{\mu\nu}q^\nu(\kappa_b^Z-i\tilde\kappa_b^Z\gamma_5)
\right] bZ^\mu \,.
\end{equation}
The influence of these couplings on \rb, as well as the asymmetry parameter
$A_b$, is presented in Fig. 1(c) from Rizzo\cite{tgr}, where the ratio of
these quantities calculated with the above Lagrangian to that of the SM
(as defined by ZFITTER\cite{zfit}) is displayed.  In this figure the solid
(dashed) curves represent the predictions when $\kappa_b^Z$
($\tilde\kappa_b^Z$)
is taken to be non-zero, with the diamonds representing unit steps of $0.01$
in these parameters.  The data points are also shown, with $m_t=170$
(dotted), 180 (solid), 190 (dashed) GeV.
Note that the present data prefer non-zero values for these couplings.

\begin{figure}[htbp]
\vspace*{10cm}
\centerline{\psfig{figure=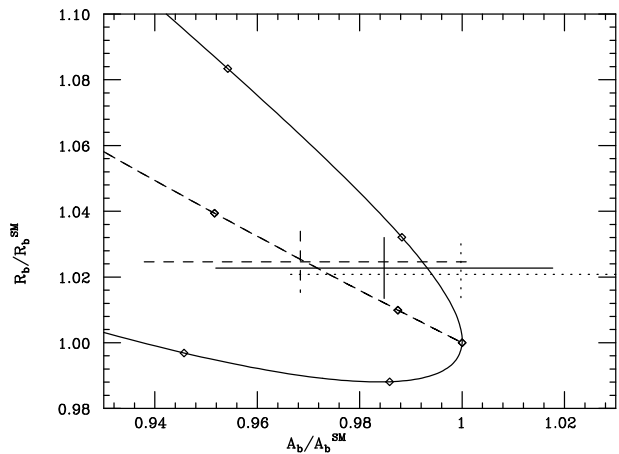,height=7.5cm,angle=90}}
\vspace*{-1cm}
\caption{\small (a) \rb\ as a function of the top-quark mass in the SM (solid
curves) and in the two-Higgs-doublet Model II (dashed curves) from Ref. 3.
The value of
the parameters are as described in the text.  The error bars indicate the
$1\sigma$ experimental measurements.  (b)  Bounds on the lightest
chargino and stop-squark masses which are consistent with \rb\ at the $1\sigma$
level from Ref. 7 for $m_t=157, 174, 191\gev$.  The allowed region lies below
the curves.  (c) The \rb\ and asymmetry parameter $A_b$ plane for non-zero
values of the electric and magnetic dipole couplings
from Rizzo in Ref. 11, where the
diamonds represent unit increments in these quantities in steps of $0.01$.
The error bars represent the data, scaled to the SM prediction with
$m_t=170, 180, 190\gev$ corresponding to the dotted, solid, dashed curves,
respectively.}
\end{figure}

\section{Radiative $B$ Decays}

Radiative $B$ decays have become one of the best testing grounds of the SM
due to recent progress on both the experimental and theoretical fronts.  The
CLEO Collaboration has recently reported\cite{cleoin} the observation of
the inclusive decay $B\to X_s\gamma$ with a branching fraction of
$(2.32\pm 0.57\pm 0.35)\times 10^{-4}$.  Observation of this process at the
inclusive level removes the uncertainties associated with folding in the
imprecisely predicted\cite{soni} ratio of exclusive to inclusive rates
when comparing theoretical results with exclusive data.  On the theoretical
side, the reliability of the calculation of the quark-level process \bsg\
is improving\cite{qcd} as agreement on the leading-logarithmic QCD corrections
has been reached and partial calculations at the next-to-leading logarithmic
order are underway.  These new results have inspired a large number of
investigations of this decay in various classes of models\cite{jlh}.

In the SM, the quark-level transition \bsg\ is mediated by $W$-boson and
t-quark exchange in an electromagnetic penguin diagram.  To obtain
the branching fraction, the inclusive rate is scaled to that of the
semi-leptonic decay $b\to X\ell\nu$.  This procedure removes uncertainties
from the overall factor of $m_b^5$,
and reduces the ambiguities involved with the imprecisely
determined Cabibbo-Kobayashi-Maskawa (CKM) factors.  The result is then
rescaled by the experimental value of $B(b\to X\ell\nu)$.
The calculation of $\Gamma(\bsg)$ employs the
renormalization group evolution\cite{qcd} for the
coefficients of the $b\to s$ transition operators in the effective Hamiltonian
at the leading logarithmic level.  The participating operators consist of
the current-current operators $O_{1,2}$, the QCD penguin operators $O_{3-6}$,
and the electro- and chromo-magnetic operators $O_{7,8}$.
The Wilson coefficients of the $b\to s$ operators are evaluated perturbatively
at the $W$ scale, where the matching conditions are imposed, and evolved
down to the renormalization scale $\mu$, usually taken to be $\sim m_b$.
This
procedure yields the branching fraction $B(\bsg)=2.92^{+0.77}_{-0.59}\times
10^{-4}$ for a top-quark mass of 175 GeV.  The central value corresponds to
$\mu=m_b$, while the upper and lower errors represent the deviation due to
assuming $\mu=m_b/2$ and $\mu=2m_b$, respectively.  We see that (i) this value
compares favorably to the recent CLEO measurement and (ii) the freedom of
choice in the value of the renormalization scale introduces an uncertainty
of order $25\%$.  Clearly, this uncertainty must be taken into account
when determining constraints on new physics from this process.

Before discussing explicit models of new physics, we first investigate the
constraints placed directly on the Wilson coefficients of the magnetic moment
operators.  Writing the coefficients at the matching scale in the form
$c_i(M_W)=c_i(M_W)_{SM}+c_i(M_W)_{new}$, where $c_i(M_W)_{new}$ represents
the contributions from new interactions, we see that the CLEO measurement
limits the possible values of $c_i(M_W)_{new}$ for $i=7,8$.  These bounds
are depicted in Fig. 2(a) for $m_t=175\gev$,
where the allowed regions lie inside the diagonal
bands.  We note that the two bands occur due to the overall sign ambiguity in
the determination of the coefficients (recall that $B(\bsg)\propto
|c_7^{eff}(\mu)|^2$), and by including the upper and lower CLEO bounds.  The
horizontal lines correspond to potential limits $B(b\to sg)<
(3-30)\times B(b\to sg)_{SM}$.
We see that such a constraint on $b\to sg$ is needed to further restrict the
values of the Wilson coefficients at the matching scale.

In two-Higgs-doublet models the
\ch\ contributes to \bsg\ via virtual
exchange together with the top-quark, and the dipole $b\to s$ operators
($O_{7,8}$) receive contributions from this exchange.  At the $W$ scale the
coefficients of these operators take the form (in Model II described above)
\begin{equation}
c_i(M_W)=G_i^{SM}(\mts/\mws)+A_{1_i}^{\ch}(\mts/\mchs)
+{1\over \tan^2\beta}A_{2_i}^{\ch}(\mts/\mchs) \,,
\end{equation}
where $i=7,8$.  The analytic
form of the functions $A_{1_i}, A_{2_i}$ can be found in Ref. 16.
In Model II, large enhancements appear for small values of \tb, but
more importantly, we see that $B(\bsg)$ is always larger than that of the SM,
independent of the value of \tb\ due to the presence of
the $A_{1_i}^{\ch}$ term.  In this case, the CLEO upper bound
excludes\cite{cleoin,me}
the region to the left and beneath the curves shown in Fig. 2(b) for
$m_t=174\pm 16\gev$ and $\mu=2m_b$.

The \ch\ couplings in
Model II are of the type present in Supersymmetry.   However, the limits
obtained in supersymmetric theories also depend on the size of the other
super-particle contributions to \bsg, and are generally much more
complex.  In particular, it has been shown\cite{bert,okada} that large
contributions can arise from stop-squark and chargino exchange (due to the
possibly large stop-squark mass splitting), as well as from the gluino and
down-type squark loops (due to left-right mixing in the sbottom sector).
The additional neutralino-down-squark contributions are expected to be small.
Some regions of the parameter space can thus cancel the \ch\ contributions
resulting in predictions for the branching fraction at (or even below) the
SM value, while other regions always enhance the amplitude.  In minimal
supergravity models with radiative breaking, the sign of the sparticle
loop contributions is found to be correlated with the sign of the higgsino mass
parameter $\mu$.
This is demonstrated in Fig. 2(c) from Ref. 19, where
$B(\bsg)$ is displayed as a function of the charged Higgs mass, for
negative and positive values of $\mu$.  The points in this figure
represent a scan of the remaining parameter space that is phenomenologically
consistent.  We see that taking $\mu<0$ $(>0)$ enhances (suppresses) the
branching fraction from the predictions in Model II.  These authors\cite{okada}
also find that $\mch>400\gev$ for $\mu<0$ and $\tb\geq 10$, while
$\mch>180\gev$ with $3\leq\tb\leq 5$ for both signs of $\mu$.

\vspace*{-0.5cm}
\nn
\begin{figure}[htbp]
\centerline{
\psfig{figure=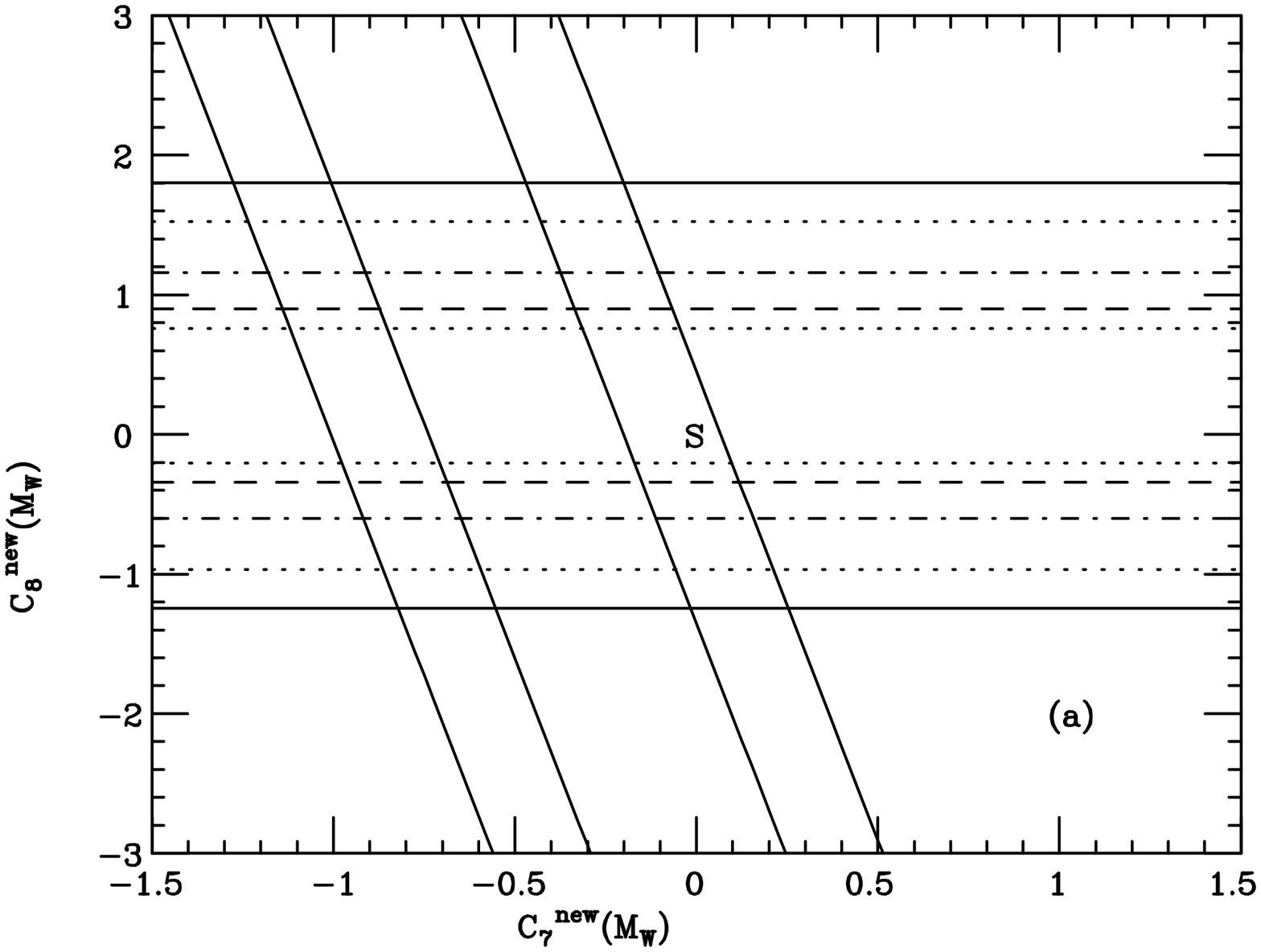,height=8.cm,width=8cm,angle=90}
\hspace*{-5mm}
\psfig{figure=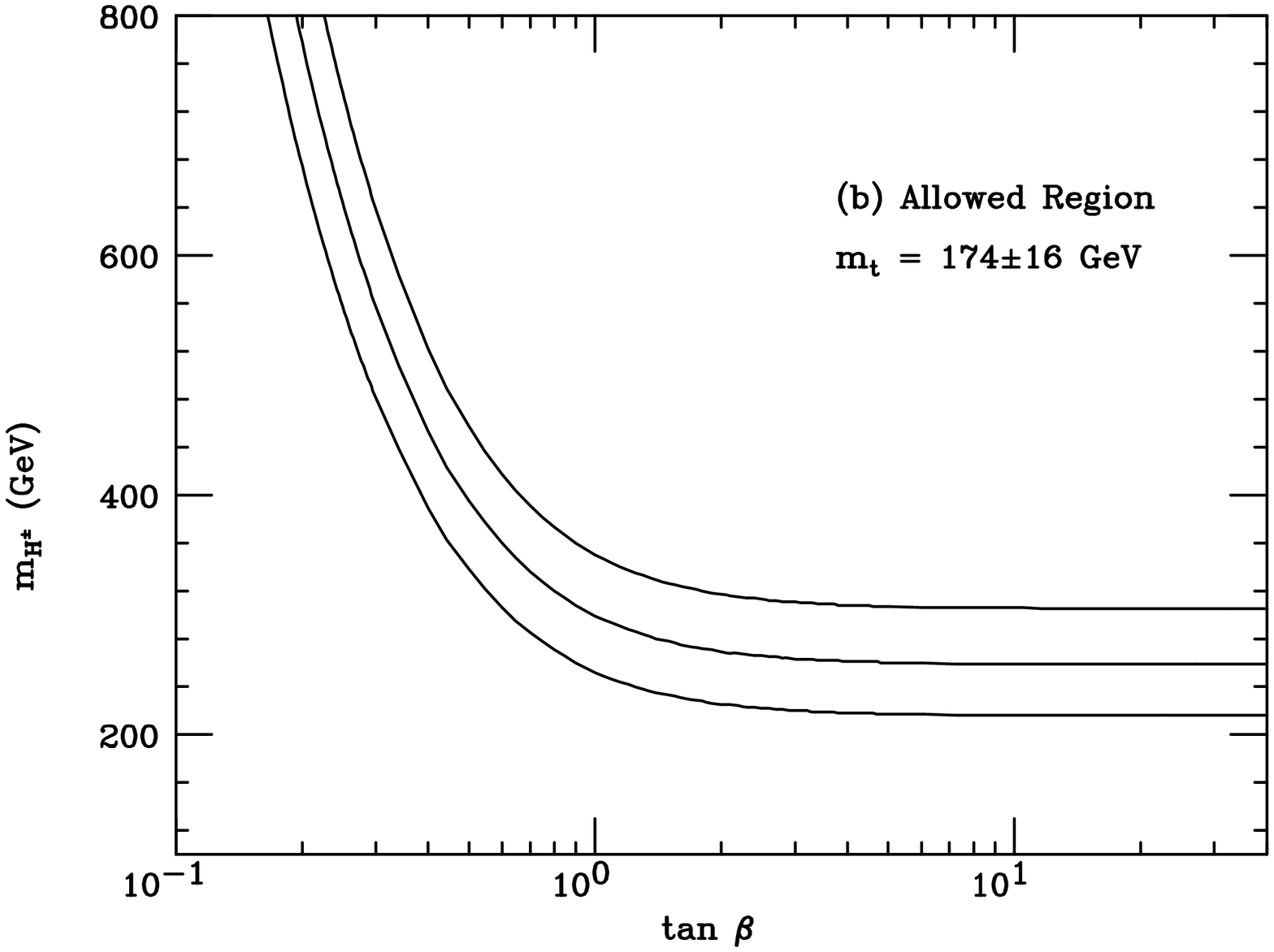,height=8.cm,width=8cm,angle=90}}
\vspace*{-0.75cm}
\centerline{
\psfig{figure=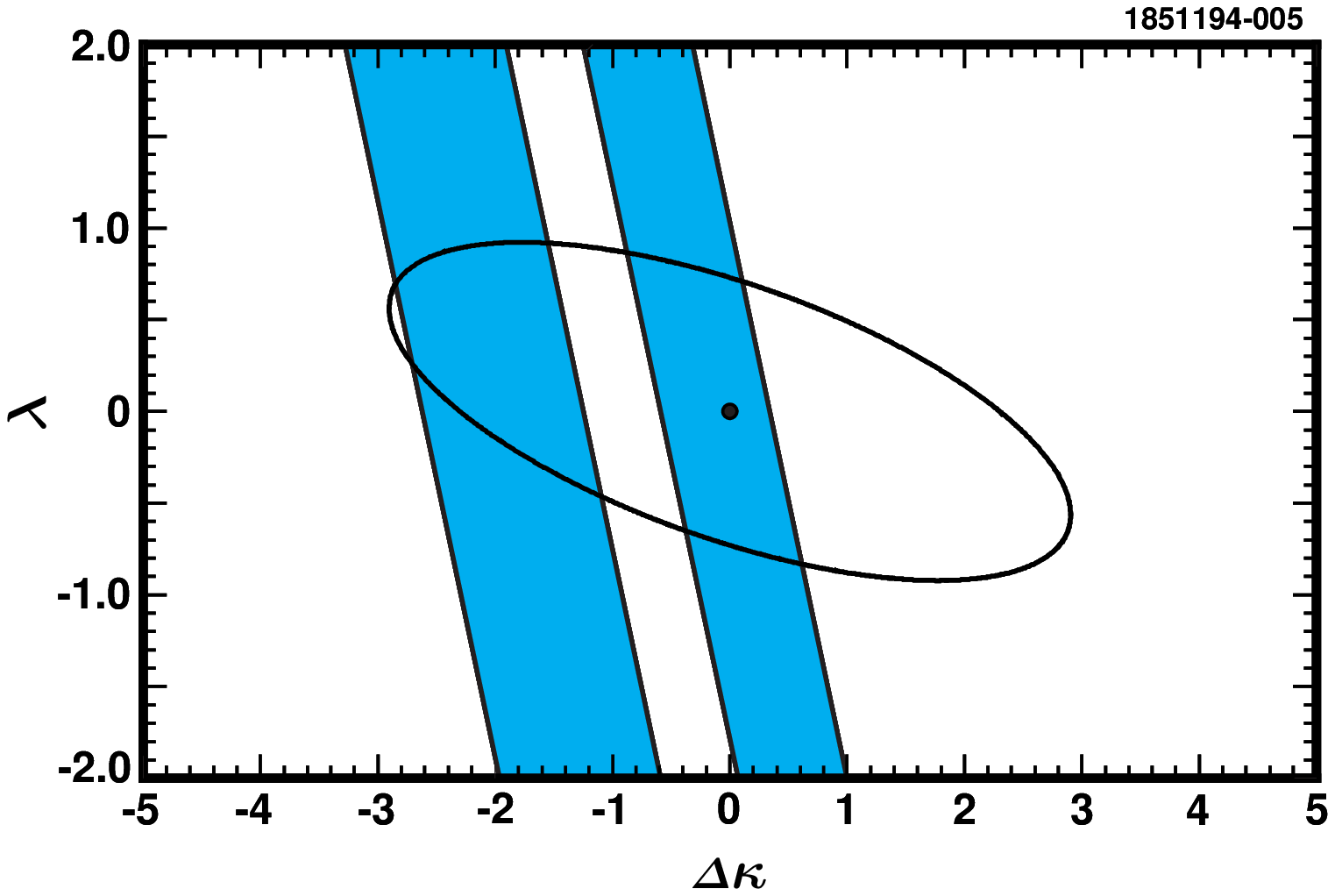,height=8.cm,width=8cm,angle=0}}
\vspace*{-1cm}
\caption{\small (a) Bounds on the contributions from new physics to $c_{7,8}$.
The region allowed by CLEO corresponds to the area inside the diagonal bands.
The horizontal lines represent potential measurements of $R\equiv
B(b\to sg)/B(b\to sg)_{SM}<30,20,10,5,3$ corresponding to the set of solid,
dotted, dash-dotted, dashed, and dotted lines, respectively.  The point
`S' represents the SM.  (b)  $B(\bsg)$ as a function of the charged Higgs
mass with $m_t=175\gev$ and $\tb=5$ from Ref. 19.  The solid curve corresponds
to the two-Higgs-doublet Model II value, while the dashed-dot curve represents
the SM.  The points represent a scan of the supersymmetric parameter space
as described in the text.
(c) Limits from \bsg\
in the charged Higgs mass - \tb\ plane.  The excluded region is that to the
left and below the curves.  The three curves correspond to the values
$m_t=190, 174, 158\gev$ from top to bottom.
(d) Constraints on anomalous $WW\gamma$ couplings from Ref. 12.  The
shaded area is that allowed by CLEO and the interior of the ellipse is the
region allowed by D0.  The dot represent the SM values.}
\end{figure}

The trilinear gauge coupling of the photon to $W^+W^-$ can also be tested
in radiative $B$ decays.  \bsg\ naturally avoids the problem of
introducing cutoffs to regulate the divergent loop integrals due to the
cancellations provided by the GIM mechanism,
and hence cutoff independent bounds on anomalous couplings can be obtained.
In this decay only the coefficient
of the magnetic dipole operator, $O_7$, is modified
by the presence of the additional terms in Eq. (1) and can be written as
\begin{equation}
c_7(M_W) = G_7^{SM}(\mts/\mws) +\Delta\kappa_\gamma A_1(\mts/\mws)
+\lambda_\gamma A_2(\mts/\mws) \,.
\end{equation}
The explicit form of the functions $A_{1,2}$ can be found in
Ref. 20.  As both of these parameters are varied, either large
enhancements or suppressions over the SM prediction for the \bsg\ branching
fraction can be obtained.  When one demands consistency with both the upper
and lower CLEO bounds, a large region of the
$\Delta\kappa_\gamma-\lambda_\gamma$ parameter plane is excluded; this
is displayed in Fig. 2(d) from Ref. 12 for $m_t=174\gev$.
Here, the allowed region is given by the cross-hatched area, where
the white strip down the middle is excluded by the
lower bound and the outer white areas are ruled out by the upper limit on
$B(\bsg)$.  The ellipse represents the region allowed by
D0\cite{dzero}.
Note that the SM point in the $\Delta\kappa_\gamma-
\lambda_\gamma$ plane (labeled by the dot) lies in the center of one of the
allowed regions.
We see that the collider constraints are complementary to those from \bsg.

\section{The Decay $B\to X_s\ell^+\ell^-$}

The inclusive process $b\to s\ell^+\ell^-$ also offers an excellent opportunity
to search for new physics.  The decay proceeds via electromagnetic and $Z$
penguin as well as by $W$ box diagrams, and hence can probe different coupling
structures than the pure electromagnetic process $b\to s\gamma$.
This reaction also receives long distance contributions
from the processes $B\rightarrow K^{(*)}\psi^{(')}$ followed by $\psi^{(')}
\rightarrow\ell^+\ell^-$ and from $c\bar c$ continuum intermediate states.
The short distance contributions  lead to the inclusive branching
fractions\cite{bsll} (including the leading logarithmic QCD corrections)
$B(B\rightarrow X_s\ell^+\ell^-)\sim (15, 7, 2)\times 10^{-6}$ for
$\ell=(e, \mu, \tau)$, respectively, and hence these modes will likely be
observed during the next few years.
The best method of separating the long and short
distance contributions, as well as observing any deviations from the
SM, is to measure the various kinematic distributions associated
with the final state lepton pair, such as the lepton pair invariant mass
distribution\cite{bsll}, the lepton pair forward-backward asymmetry\cite{ali},
and the tau polarization asymmetry\cite{mebsll} in the case $\ell=\tau$.
Measurement of all these quantities would allow for the determination of the
sign and magnitude of the Wilson coefficients for the electroweak loop
operators and thus provide a completely model independent analysis.
We note that measurement of these distributions requires the high
statistics samples which will be available at future B-factories.
The lepton pair invariant mass distribution for $b\to se^+e^-$ is displayed
in Fig. 3(a) (taking $m_t=175\gev$),
where the solid curve includes the contributions from the short
and long range effects and the dashed curve represents the short distance
alone.
We see that the long distance contributions
dominate only in the $M_{e^+e^-}$ regions near the $\psi$ and $\psi'$
resonances, and observations away from these peaks would cleanly separate the
short distance physics.  The tau polarization asymmetry is
presented in Fig. 3(b); we see that it is large and negative for this value
of $m_t$.  As an example of how new physics can affect this process, we
examine $b\to s\ell^+\ell^-$ in the case of an anomalous $WW\gamma$ vertex.
The resulting invariant mass spectrum is shown in Fig. 3(c) for several
values of $\Delta\kappa_\gamma$ (taking $\lambda_\gamma=0$), and the
variation of the tau polarization asymmetry with non-zero values of
$\Delta\kappa_\gamma$ and $\lambda_\gamma$ is displayed in Fig. 3(d)
for $\hat s\equiv q^2/m^2_b=0.7$.

\vspace*{-0.5cm}
\nn
\begin{figure}[htbp]
\centerline{
\psfig{figure=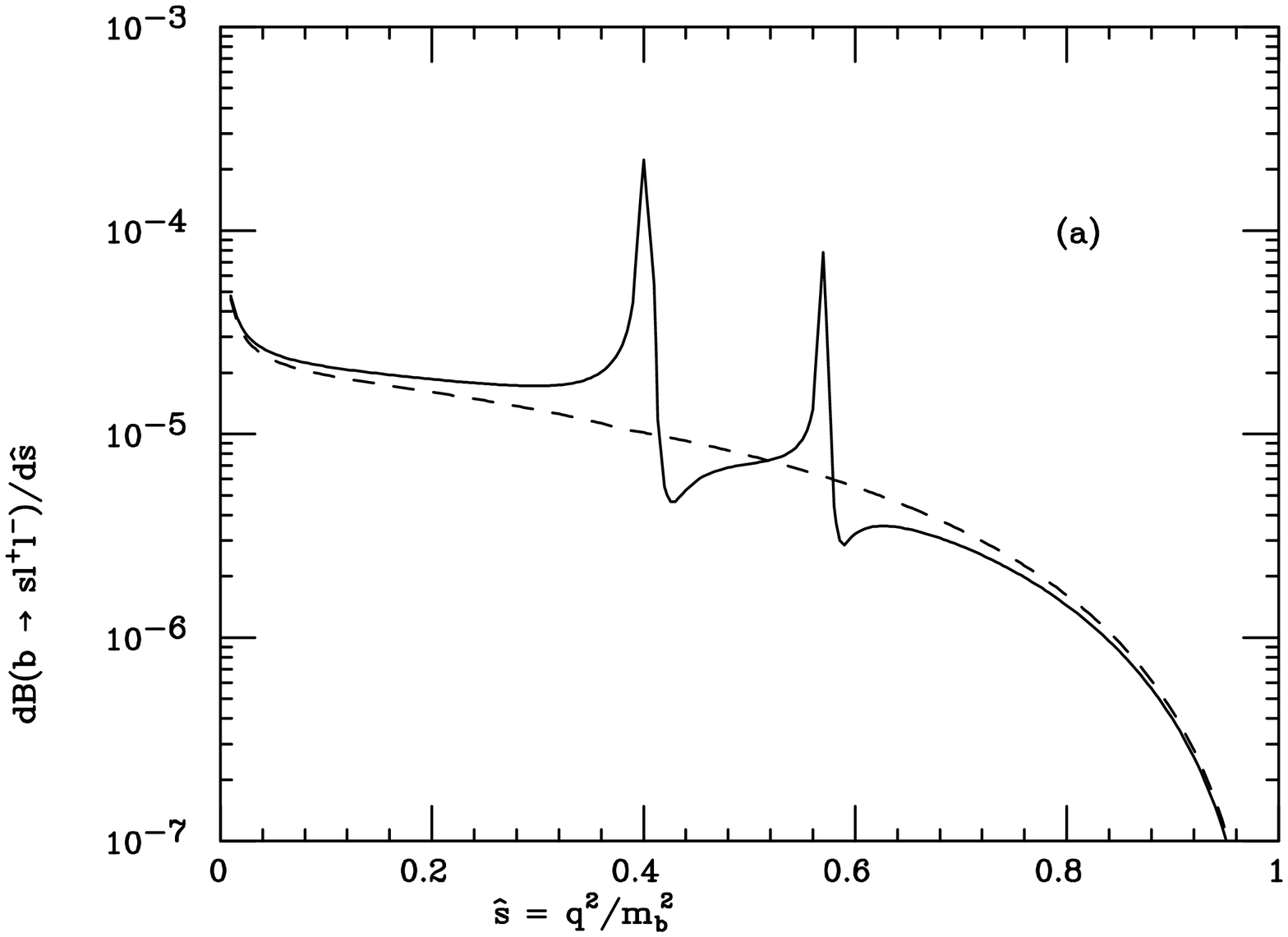,height=8.cm,width=8cm,angle=90}
\hspace*{-5mm}
\psfig{figure=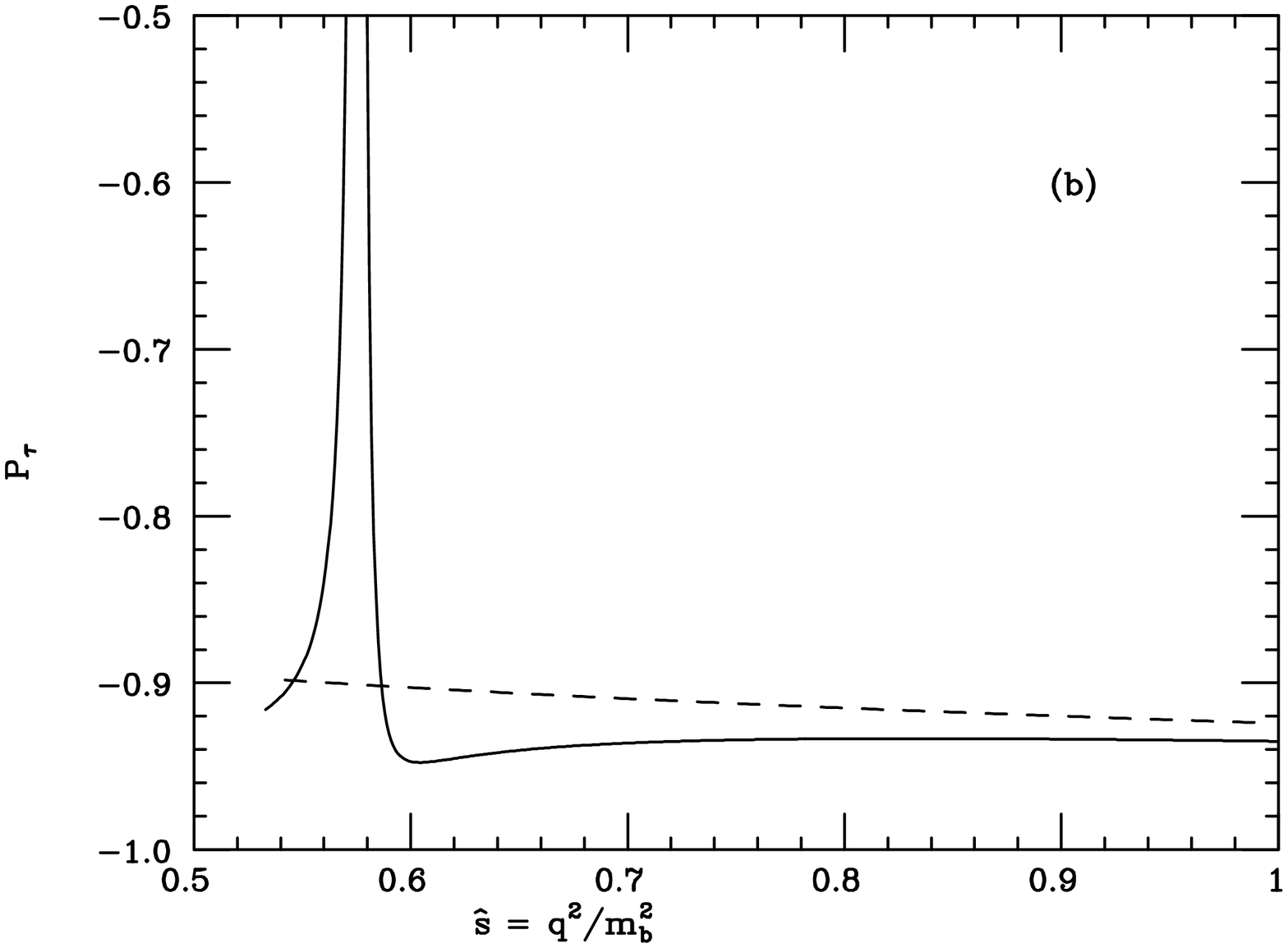,height=8.cm,width=8cm,angle=90}}
\vspace*{-0.75cm}
\centerline{
\psfig{figure=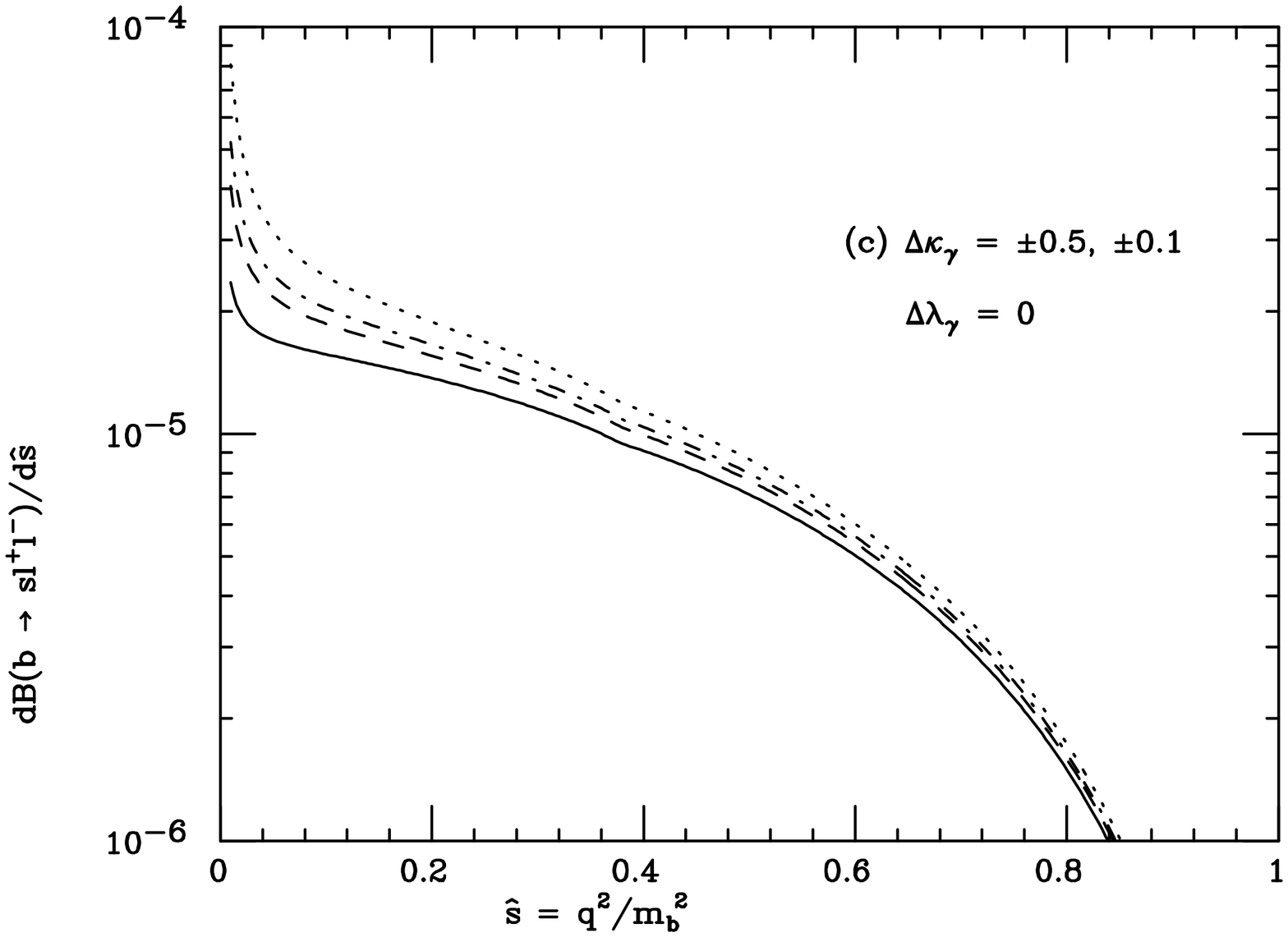,height=8.cm,width=8cm,angle=90}
\hspace*{-5mm}
\psfig{figure=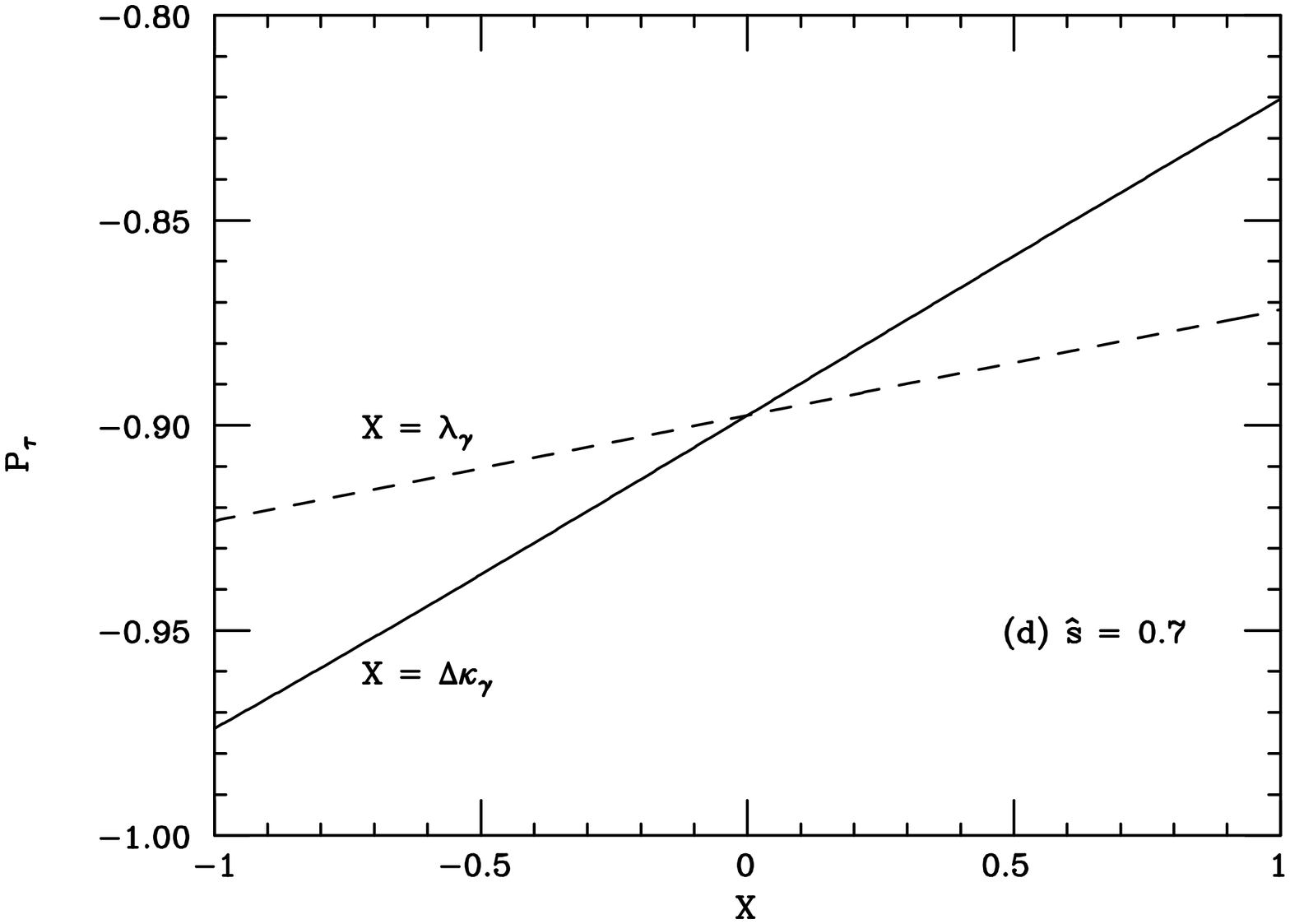,height=8.cm,width=8cm,angle=90}}
\vspace*{-1cm}
\caption{\small The (a) lepton pair mass distribution (with $\ell=e$) and (b)
tau polarization asymmetry (with $\ell=\tau$) in the SM, and the (c) lepton
pair mass distribution
and (d) tau polarization asymmetry with anomalous $WW\gamma$ couplings as
labeled, for the process $b\to s\ell^+\ell^-$ with $m_t=175\gev$.}
\end{figure}

\section{CP Violation in $B$ Decays}

CP violation in the $B$ system will be examined\cite{babar} during the next
decade at dedicated B-Factories.  CP violation arises in the SM from the
existence of the phase in the 3 generation CKM matrix.  The
relation $V_{tb}V^*_{td}+V_{cb}V^*_{cd}+V_{ub}V^*_{ud}=0$, which is required
by unitarity, can be depicted as a triangle in the complex plane, where
the area of the triangle represents the amount of CP violation.
It can be shown that the apex of the triangle is located at the
point $(\rho,\eta)$ in the complex plane, where $\rho$ and $\eta$ are
parameters describing the CKM matrix in the Wolfenstein notation.
The present status of these parameters is summarized in Fig. 4(a),
where the shaded area is that allowed in the SM.
This region is determined by measurements
of the quantities (i) $|V_{ub}|$ and $|V_{cb}|$, (ii) $\epsilon$, the CP
violation parameter in $K^0_L$ decay, and (iii) the
rate for $B^0_d-\bar B^0_d$ mixing,
together with theoretical estimates for the parameters which relate these
measurements to the underlying theory, such as $B_K,\ f_B,$ and $B_B$.
The value of $\overline{m_t}(m_t)$ is taken to be consistent with
the physical range $174\pm 16\gev$.  This yields the
allowed ranges for the angles of the triangle: $-0.89\leq\sin 2\alpha\leq
1.00,\
0.18\leq\sin 2\beta\leq 0.81$, and $-1.00\leq\sin 2\gamma\leq 1.00$.

\vspace*{-0.5cm}
\nn
\begin{figure}[htbp]
\centerline{
\psfig{figure=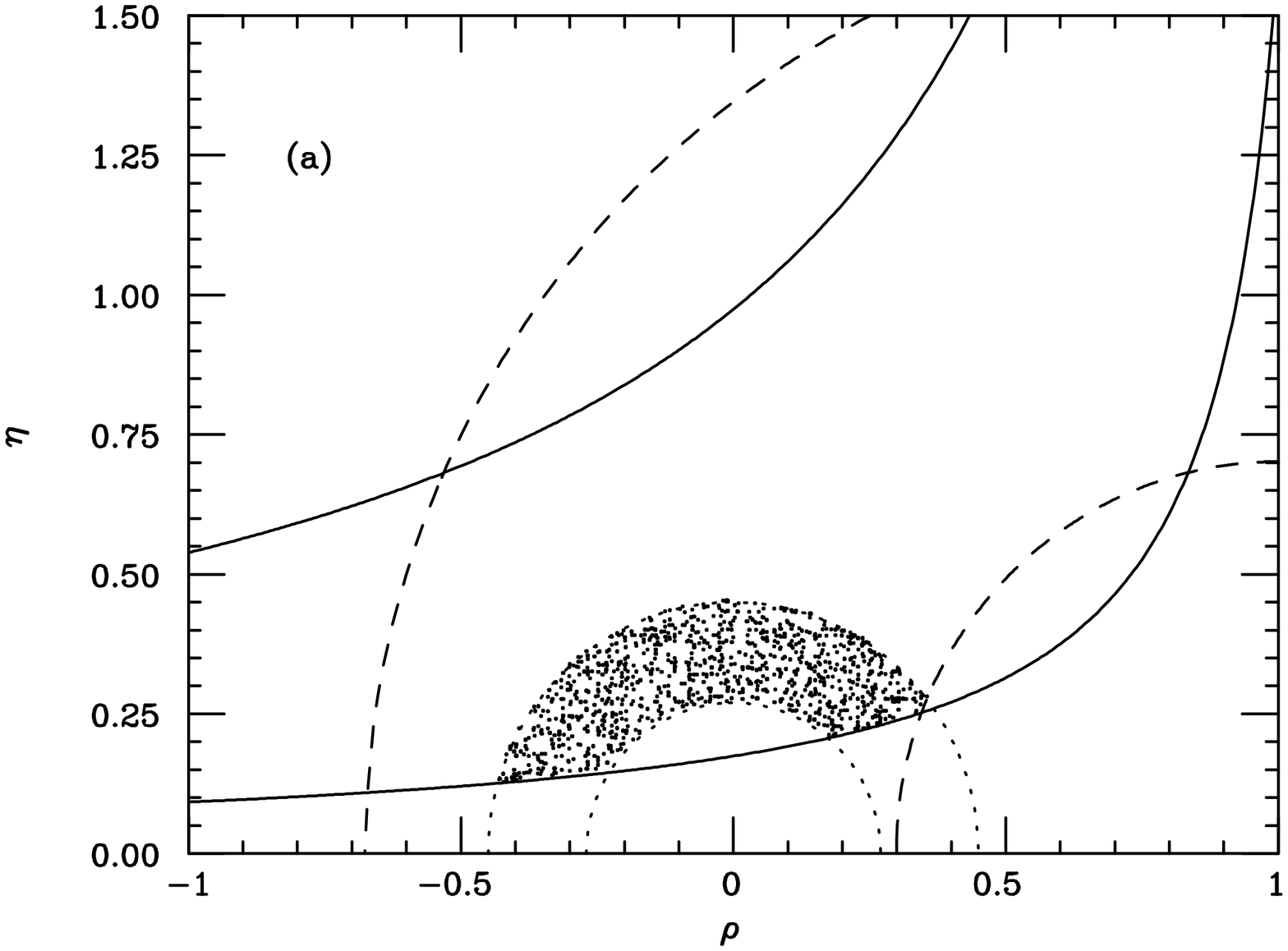,height=8cm,width=8cm,angle=90}
\hspace*{-5mm}
\psfig{figure=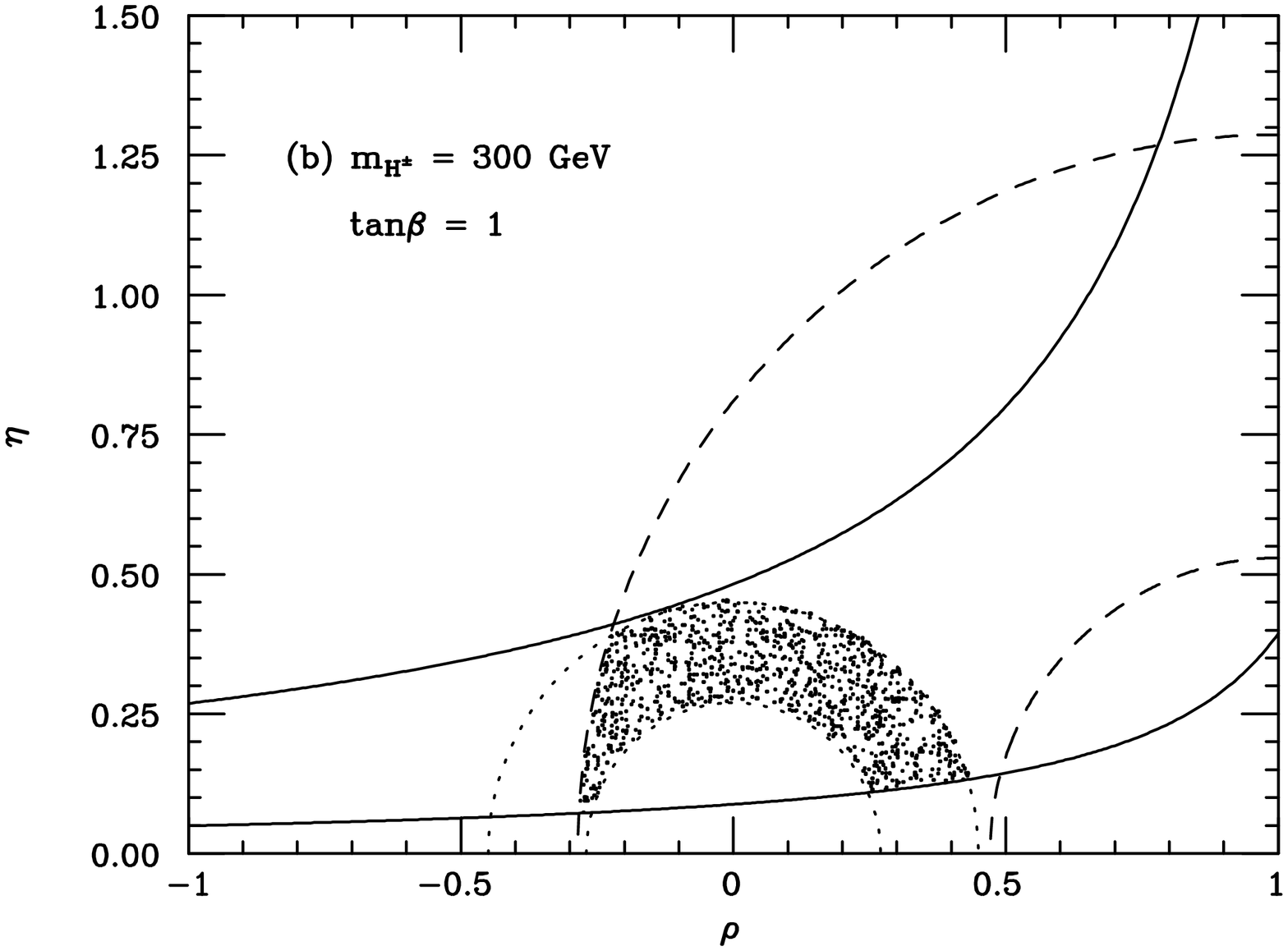,height=8cm,width=8cm,angle=90}}
\vspace*{-1cm}
\caption{\small Constraints in the (a) SM and (b) two-Higgs-doublet Model II in
the $\rho-\eta$ plane from $|V_{ub}|/|V_{cb}|$ (dotted circles), $B_d^0-
\bar B_d^0$ mixing (dashed circles) and $\epsilon$ (solid hyperbolas).
The shaded area corresponds to that allowed for the
apex of the Unitarity triangle.}
\end{figure}

It is important to remember that this picture can be dramatically altered if
new physics is present, even if there are no new sources of CP violation.
Figure 4(b) displays the constraints in the $\rho-\eta$ plane
in the two-Higgs-doublet Model II.  In this case the presence of the
extra Higgs doublet is felt by the virtual exchange of the \ch\ boson in
the box diagram which mediates $B^0_d-\bar B^0_d$ mixing and governs the
value of $\epsilon$.  For this $\rho-\eta$ region, the allowed ranges of
the angles of the unitarity triangle become $-1.00\leq\sin 2\alpha\leq 1.00,\
0.12\leq\sin 2\beta\leq 0.81$, and $-1.00\leq\sin 2\gamma\leq 1.00$.
We see that the SM predictions for CP violation are thus modified.

\section{Summary}

We have examined several aspects of physics in the $B$ system in
a variety of models containing physics beyond the SM and discovered that
these processes can provide powerful insights on new
interactions.  In some cases, such as in \bsg, constraints are obtained which
either complement or are stronger than those from other
low-energy processes or from direct collider searches.  The decay $b\to
s\ell^+\ell^-$ is also an excellent probe of new physics and we eagerly
anticipate its detection.  We also await further improvements in the data on
$Z\to b\bar b$ to see if this is the process which finally cracks the SM.
In summary, we have an exciting decade of $B$ physics ahead!

\vspace{1.0cm}

%
\def\MPL #1 #2 #3 {Mod.~Phys.~Lett.~{\bf#1},\ #2 (#3)}
\def\NPB #1 #2 #3 {Nucl.~Phys.~{\bf#1},\ #2 (#3)}
\def\PLB #1 #2 #3 {Phys.~Lett.~{\bf#1},\ #2 (#3)}
\def\PR #1 #2 #3 {Phys.~Rep.~{\bf#1},\ #2 (#3)}
\def\PRD #1 #2 #3 {Phys.~Rev.~{\bf#1},\ #2 (#3)}
\def\PRL #1 #2 #3 {Phys.~Rev.~Lett.~{\bf#1},\ #2 (#3)}
\def\RMP #1 #2 #3 {Rev.~Mod.~Phys.~{\bf#1},\ #2 (#3)}
\def\ZP #1 #2 #3 {Z.~Phys.~{\bf#1},\ #2 (#3)}
\bibliographystyle{unsrt}

\begin{thebibliography}{99}
%
\bibitem{schaile}
D.\ Schaile, in {\it 27th International Conference on High Energy Physics},
Glasgow, Scotland, July 1994; U. Uwer in {\it 30th Rencontres de Moriond:
Electroweak Interactions and Unified Theories}, Meribel les Allures, France,
March 1995.
%
\bibitem{cdfdo}
F. Abe \etal, (CDF Collaboration), FERMILAB-PUB-95-022-E (1995);
S. Abachi \etal, (D0 Collaboration), FERMILAB-PUB-95-028-E (1995).
%
\bibitem{grant}
A.K. Grant, \PRD D51 207 1995 .
%
\bibitem{zfit}
The ZFITTER package: D.\ Bardin \etal, \ZP C44 493 1989 ; \NPB B351 1 1991 ;
\PLB B255 290 1991 ; CERN Report, CERN-TH-6443/92, 1992.
%
\bibitem{hhg}
J.F. Gunion, H.E. Haber, G.L. Kane, and S. Dawson, {\it The Higgs Hunters
Guide}, (Addison-Wesley, Redwood City, CA, 1990).
%
\bibitem{zbbsusy}
A. Djouadi \etal, \NPB B349 48 1991 ;
A. Denner \etal, \ZP C51 695 1991 ;
M. Boulware and D. Finnel, \PRD D44 2054 1991 .
%
\bibitem{wkk}
J.D. Wells, C. Kolda, and G.L. Kane, \PLB B338 219 1994 .
%
\bibitem{burgess}
C.P. Burgess and D. London, Phys. Rev. Lett. {\bf 69}, 3428 (1992).
%
\bibitem{dieter}
K. Hagiwara \etal, \NPB B282 253 1987 .
%
\bibitem{eboli}
O.J.P. Eboli, M.C. Gonzalez-Garcia, and S.F. Novaes, \PLB B339 119 1994 .
%
\bibitem{tgr}
T.G. Rizzo, \PRD D51 3811 1995 ; G. Kopp \etal, \ZP C65 545 1995 .
%
\bibitem{cleoin}
M.S. Alam \etal, (CLEO Collaboration), \PRL 74 2885 1995 .
%
\bibitem{soni}
C.\ Bernard, P.\ Hsieh, and A.\ Soni, Phys. Rev. Lett. {\bf 72}, 1402 (1994).
%
\bibitem{qcd}
A.J. Buras \etal, \NPB B424 374 1994 , and references therein.
%
\bibitem{jlh}
For a review of implications of non-SM physics in \bsg, see,
J.L. Hewett, in {\it 21st Annual SLAC Summer Institute
on Particle Physics}, Stanford, Ca, July 1993.
%
\bibitem{bsgch}
T.G.\ Rizzo,
Phys.\ Rev.\ {\bf D38}, 820 (1988); W.-S.\ Hou and R.S.\ Willey,
Phys.\ Lett.\ {\bf B202}, 591 (1988); C.Q.\ Geng and J.N.\ Ng,
Phys.\ Rev.\ {\bf D38}, 2858 (1988);
B.\ Grinstein, R.\ Springer, and M.\ Wise,
Nucl.\ Phys.\ {\bf B339}, 269 (1990);
V.\ Barger, J.L.\ Hewett, and R.J.N.\ Phillips,
Phys.\ Rev.\ {\bf D41}, 3421 (1990).
%
\bibitem{me}
J.L. Hewett, Phys. Rev. Lett. {\bf 70}, 1045 (1993);
V. Barger, M. Berger, and R.J.N. Phillips, Phys. Rev. Lett. {\bf 70},
1368 (1993).
%
\bibitem{bert}
S.\ Bertolini, {\it et al.},
Nucl.\ Phys.\ {\bf B294}, 321 (1987), and Nucl.\ Phys.\ {\bf B353}, 591 (1991);
R. Barbiero and G.F. Giudice, \PLB B309 86 1993 ; R. Garisto and J.N. Ng,
\PLB B315 119 1993 ; M.A. Diaz, \PLB B322 207 1994 ; F.M. Borzumati
\ZP C63 291 1994 ; S. Bertolini and F. Vissani, Trieste Report SISSA 40/94/EP;
J.L. Lopez \etal, \PRD D48 974 1993 .
%
\bibitem{okada}
T. Goto and Y. Okada, KEK Report KEK-TH-421 (1994).
%
\bibitem{tgrtwo}
T.G. Rizzo, Phys. Lett. {\bf B315}, 471 (1993); S.-P. Chia, Phys. Lett.
{\bf B240}, 465 (1990); K.A. Peterson, Phys. Lett. {\bf B282}, 207 (1992).
%
\bibitem{dzero}
J. Ellison (D0 Collaboration), Proceeding of {\it 1994 Meeting of the Division
of Particles and Fields}, Albuquerque, NM (1994).
For comparable bounds from CDF and UA2, see
F. Abe \etal, (CDF Collaboration) \PRL 74 1936 1995 ;
J. Alitti \etal, UA2 Collaboration, Phys. Lett. {\bf B277}, 194 (1992).
%
\bibitem{bsll}
N.G. Deshpande and J. Trampetic, \PRL 60 2583 1988 ; C.S. Lim, T. Morozumi,
and A.I. Sanda, \PLB B218 343 1989 ; N.G. Deshpande, J. Trampetic, and
K. Panrose, \PRD D39 1461 1989 ; B. Grinstein, M.J. Savage, and M.B. Wise,
\NPB B319 271 1989 .
%
\bibitem{ali}
A. Ali, G.F. Guidice, and T. Mannel CERN Report CERN-TH.7346/94; A. Ali,
T. Mannel, and T. Morozumi, \PLB B273 505 1991 .
%
\bibitem{mebsll}
J.L. Hewett, SLAC Report, SLAC-PUB-95-6820.
%
\bibitem{babar}
D. Boutigny \etal, SLAC Report SLAC-0443 (1994); M.T. Cheng \etal,
KEK Report KEK-94-02 (1994).
%

\end{thebibliography}

\end{document}